# Comprehensive G-C Modeling of a Gapless Continuously Variable Series Reactor

Mohammadali Hayerikhiyavi, Aleksandar Dimitrovski
Department of Electrical and Computer Engineering
University of Central Florida, Orlando, USA
Mohammad.ali.hayeri@knights.ucf.edu, ADimitrovski@ ucf.edu

*Abstract--* The paper presents a comprehensive model of Continuously Variable Series Reactor (CVSR) in the Gyrator-Capacitor (G-C) framework. CVSR is a device capable of controlling the reactance in an ac circuit using the nonlinear characteristic of ferromagnetic core. Bias magnetic flux produced by a dc winding can be used to regulate the equivalent ac reactance for various applications, such as power flow control, oscillation damping, or fault current limiting. A novel variant of CVSR that features gapless ferromagnetic core and a split two-coil ac winding is considered. Its nominal reactance is entirely determined by leakage fluxes. The Gyrator-Capacitor (G-C) approach is used to model the interface between the magnetic, electronic, and electric circuits, to maintain consistency of active and reactive powers in these circuits and allow additional flexibility. For accurate representation of the gapless CVSR, in addition to a realistic nonlinear and lossy core, elaborate models of different flux leakages are given. Results from analyses for a range of distances between the coils of the ac winding at different bias currents are presented.

## I. Introduction

Reactors are one of the fundamental passive elements in power systems that are used to provide economical and efficient solutions to some common operational problems. Most frequently, they are used for reactive power management and fault current limiting. Variable reactors could also be used, if they existed at required high power levels, to damp transients and control power flows. Devices traditionally used for these purposes include switchable capacitors/reactors, generator controls, and phase-shifting transformers. In the past few decades, various types of power electronics-based flexible ac transmission systems controllers (FACTS) have been developed to complement them [1]. However, all of the aforementioned devices can either provide only a limited functionality or come at a very high cost. More recently, continuously variable series reactor (CVSR) technology based on the magnetic amplifier principle [2] has been proposed as an alternative option with high reliability and low costs [3]. In essence, it is a hybrid technology that uses traditional robust electromagnetic components with a low-rated power electronics-based control circuit. This is enabled by galvanic isolation of the power and the control circuit, with their interaction through the magnetic field only.

In principle, variable reactors can be used in both shunt and series connection. As the name implies, the focus here is on the latter. The ac winding of the CVSR is connected in series with the power line to insert variable reactance within the device design boundaries. The continuous control of this reactance directly impacts the current and power flow in the power line and allows for continuous control of steady state power flows, In addition, with fast controls provided by the power electronics, it can be used in other dynamic applications such as oscillation damping [4] and, with some modifications, fault current limiting. The broad range of applications, of course, requires studies of the mutual impacts of the CVSR and the power system during all potential conditions.

The Gyrator-Capacitor (G-C) approach is a very effective method for modeling electromagnetic circuits for detailed analysis of power magnetic devices [2, 5, 6]. It seamlessly links electrical and magnetic domains for a comprehensive study of the complex, realistic hybrid devices installed in the power system. With this approach, the analogy between the magneto-motive force (MMF) and the electromotive force (voltage source) is as usual, but the electrical current represents the rate of change of the magnetic flux instead of the flux itself. Hence, magnetic permeance (inverse magnetic reluctance) becomes analogous to capacitance instead of conductance. This allows for correct representation of the interchanged energy in a hybrid electromagnetic device such is the CVSR [3].

Accurate models of complex devices used in integrated power system simulations are fundamental to enhance understanding of the behavior of both the elements and power system. Frequently, for example, magnetic cores of power system devices are modeled in a simplistic fashion. But, for a device like CVSR whose magnetic characteristics are essential for its performance, a much more accurate model is needed [7]. For a gapless CVSR, in particular, flux leakages significantly affect the magnetic circuit behavior, and their inclusion is necessary to achieve good match between the model and the real device. Not incorporating the effects of different leakages has been shown to be the dominant error in the analysis. The improvements to the existing model of a gapless reactor to reduce the discrepancies, in particular in presence of a bias flux, are the main contribution of this work.

The rest of the paper is structured as follows. The basic concept of the CVSR is briefly reviewed in Section 2; the G-C modeling approach is described in Section 3; Section 4 describes a comprehensive G-C model of the single-phase gapless CVSR; Section 5 presents results from a case study, and conclusions are drawn in Section 6.

## II. Gapless CVSR

For the single-phase CVSR presented in Figure 1, two coils of the ac winding are wound on the middle leg of a three-legged magnetic circuit, and connected in series with the ac power circuit. The two coils are connected in a counter-series orientation with respect to each other, i.e. they create magnetic fluxes that oppose each other. This results with a virtual gap in the middle between the coils that extends through the core windows, as illustrated by the ac flux lines in Figure 1. Due to the use of the window space, the saturation of the ferromagnetic core is prevented even at large load currents, despite the fact that there is no air gap in the core itself. The dc winding also consists of two coils that are wound on the outer legs and connected in series. This winding is connected to a dc source that controls the bias dc current and the resulting bias dc flux.

The nominal reactance of the CVSR, defined by the self-inductance of the ac winding, is the reactance without any bias flux in the circuit. It is determined by the distance between the two coils in the ac winding (DBW) and, it can be shown, that its value is the largest when the two ac coils are farthest apart, and the smallest when the two coils are next to each other. This reactance is further regulated by the bias dc current. The largest value is the nominal value when the core is in the linear region of the B-H curve (at no or very small dc) and the smallest value is when the outer parts of the core are entirely saturated (at large dc) [2]. Consequently, the equivalent reactance of the ac circuit is impacted by the two parameters, $I_{dc}$ and DBW. Theoretically, DBW can be negative, meaning the two ac coils overlap each other partially or completely. The latter is the limiting case which will result with zero nominal inductance. In practice, of course, this cannot be accomplished nor is a need for it.

At any time, the fluxes passing through the outer legs are, in general, combination of the ac and the dc flux. They will add in one of the outer legs and subtract in the other, as shown in Figure 1. Hence, the induced voltages in the right and the left dc coil will be $\frac{d\Phi_{right}}{dt}$ and $\frac{d\Phi_{left}}{dt}$, respectively, and the induced voltage across the entire dc winding is $V_{bias} = V_{right} - V_{left}$.

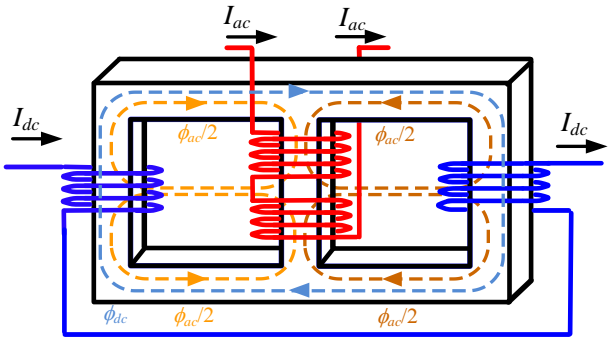

Fig. 1. Simplified schematic of a gapless CVSR [1].

### III. GYRATOR-CAPACITOR EQUIVALENT

A general approach to modelling magnetic circuits is to use electric circuit analogy. Equivalent circuits are typically formed using resistors that represent flux paths, and voltage sources that represent MMFs [8]. However, there is an inconsistency in using dissipative elements like resistors to represent energy storage elements like magnetic cores. An energy equivalence approach is presented by the G-C model as illustrated in Figure 2 [9, 10].

In this approach, the analogy between MMF and voltage source stays the same, but current represents rate-of-change of magnetic flux $\frac{d\Phi}{dt}$, as described by (1) and (2). The gyrator acts as a dualizer that allows voltage and current interchange based on the number of turns $N$ in the winding, as described by (3) and (4), which defines the gyrator resistance $G$ by (5).

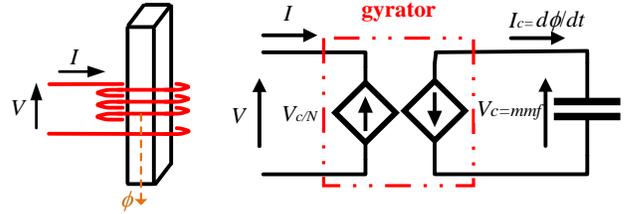

Fig. 2. An electromagnetic circuit and its equivalent gyrator-capacitor model

$$V_c \equiv mmf = RI \tag{1}$$

$$V = -RI_c = -R\frac{d\Phi}{dt} \tag{2}$$

$$I = \frac{mmf}{N} \tag{3}$$

$$V = -N\frac{d\Phi}{dt} \tag{4}$$

$$G = \frac{I}{V} = \frac{1}{N} \tag{5}$$

These expressions lead to an equivalent representation of magnetic permeances (inverse reluctances) with capacitances.

The nonlinear magnetic paths are well defined by the geometry of the ferromagnetic circuit. The corresponding nonlinear reluctances can simply be calculated from the B-H characteristic of the ferromagnetic material and the geometric parameters. They can be expressed approximately with (6):

$$\mathcal{R}(\Phi) = \frac{\mu_r \mu_0 l}{A} \tag{6}$$

where:
$\mu_0 = 4\pi \times 10^{-7}$ – magnetic permeability of free space,
$\mu_r$ – relative magnetic permeability of the material,
$A$ – cross-sectional area,
$l$ – mean length of the path.

The nonlinear permeances which are reciprocal values of these reluctances are modelled as nonlinear capacitors. The other leakage permeances are much more complicated to calculate and they are discussed in the following section.

### IV. G-C MODEL OF A GAPLESS CVSR

A comprehensive G-C model of the gapless CVSR with all of its components is shown in Figure 3. As discussed above, the ferromagnetic core is represented with nonlinear permeances. The nonlinear permeances representing different nonlinear magnetic paths in the CVSR are modeled with variable capacitors ($C_1 - C_{20}$), as calculated by (6). The other linear permeances represent the different flux leakage paths.

A flux leakage in a magnetic circuit is the flux outside of the intended path in the circuit. It is the result of a relatively small (order of thousands) permeability difference between the ferromagnetic core and the non-ferromagnetic environment. Usually, flux leakages are represented by lumped equivalent reactances that are determined by measurements as, for example, short circuit test of a transformer. This is not viable in a device such as CVSR where individual leakage components are necessary to construct an accurate model. Hence, they need to be derived analytically.

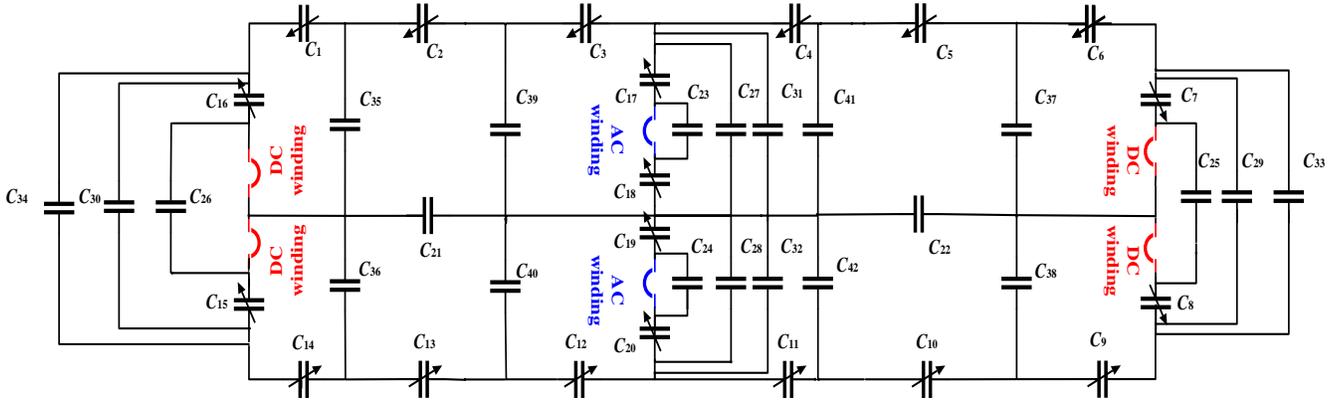

Fig. 3. Comprehensive Gyrator-Capacitor model of a gapless CVSR

Flux leakage components and corresponding leakage permeances can be grouped into the following categories: 2D slot leakage, 3D slot leakage, and exterior adjacent leakage, as shown for the simple magnetic circuit in Figure 4 [7]. In order to show the flux paths clearly, the core is made transparent.

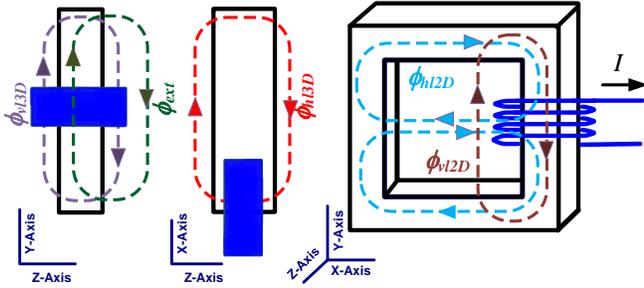

Fig. 4. Flux leakages

The 2D leakage flux is inside the core window and passes across the window slot in the X-Y plane. The slot leakage flux is broken down into horizontal and vertical components corresponding to the horizontal and vertical components of the magnetic field intensity. These leakage fluxes curve over the front and the back face as well. Hence, the corresponding 3D slot leakage permeances should be added to the 2D slot leakage permeances for a more accurate model. The vertical and horizontal slot leakage paths on the front and the back face (3D slot leakages) are shown on the two side views on the left in Figure 4. Another leakage flux that should be taken into account is the exterior adjacent leakage flux. It is the flux associated with the coil bundle adjacent to the core, external to the slot, on the front and the back of the leg ($\emptyset_{ext}$), as shown in the leftmost Y-Z plane in Figure 4.

As described above, these leakage permeances lead to the following groups of capacitances in the G-C model:
- $C_{23} - C_{26}$: horizontal leakage permeances
- $C_{27} - C_{30}$: vertical leakage permeances inside coils
- $C_{31} - C_{34}$: exterior leakage permeances
- $C_{35} - C_{42}$: vertical leakage permeances outside coils

In calculating the leakage permeances, the following two approximations are made:

a) There are no significant MMF drops across all the paths within the core, because the permeability of the ferromagnetic core is much higher than that of the surround.
b) The magnetic field intensity is constant along the given leakage path.

The permeances are related to the energy stored in the magnetic field. In a linear inductor, the energy can be expressed in terms of the coil current, the number of coil turns, and the leakage permeance as:

$$E = \tfrac{1}{2} P N^2 i^2 \qquad (7)$$

On the other hand, the energy stored in a linear inductor from the field perspective is given with:

$$E = \tfrac{1}{2}\mu_0 \int H^2 dV \qquad (8)$$

where $H$ is magnetic field intensity, and $dV$ is volumetric integration variable over the volume of the coil window. By decomposing the field intensity into orthogonal components which are not mutually coupled, separating the space inside and outside of the winding, and applying Ampere's law, the corresponding leakage permeances can be obtained using (7) and (8). Expressions for all of them are provided in Appendix.

## V. CASE STUDY

In order to illustrate the model and its performance, a gapless CVSR with cross-section geometry as shown in Fig. 5 is considered. The core windows have width $w_s$ and height $d_s$, while the core has depth $l_c$ that extends into the page in the third dimension. The widths of the central and the outer legs are $w_c$ and $w_o$, respectively, and the height of the yoke is $d_b$. The bundle of conductors in the ac coils has width $w_w$ and depth $d_w$, and the distance between the coils is denoted as DBW. For simplicity, it is assumed that all of the coils have rectangular profile. Symbols ⊗ and ⊙ indicate directions of the current in and out of the page, respectively. The model from Fig. 3 was implemented in Simulink®, and the CVSR was connected in series in a circuit with a source and a load impedance. The numerical values for all of the CVSR parameters and the test circuit are given in Table I.

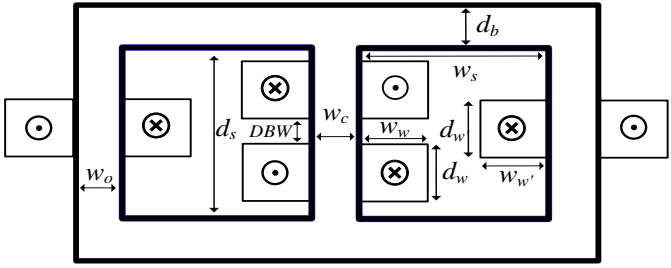

Fig. 5. Gapless CVSR cross-section geometry

Table I – GAPLESS CVSR PARAMETERS

| Parameter | Description | Value |
|---|---|---|
| $d_s$ | Window height | 24 inch |
| $w_s$ | Window width | 4 inch |
| $d_w$ | Height of ac coil | 5 inch |
| $w_w$ | Width of ac coil | 0.6 inch |
| $d_w'$ | Height of dc coil | 5 inch |
| $w_w'$ | Width of dc coil | 0.6 inch |
| $l_c$ | Core depth | 4 inch |
| $w_c$ | Outer legs width | 8 inch |
| $w_0$ | Middle leg width | 4 inch |
| $d_b$ | Yoke and base height | 4 inch |
| $N_{dc}$ | Number of turns in dc winding | 450 |
| $N_{ac}$ | Number of turns in ac winding | 300 |
| $V$ | Ac rated voltage | 2.4 kV (RMS) |
| $I$ | Ac rated current | 20.9 A (RMS) |
| $R$ | Load resistance | 100 Ω |
| $L$ | Load inductance | 130 mH |
| $PF$ | Power factor | 0.9 |
| $S$ | Rated power | 50 kVA |
| $B_{sat}$ – M36 | Saturation point | 1.34 T |
| DBW | Distance between coils | 12 to -5 inch |

Two variants of the M36 material B-H characteristic are shown in Figure 6: (a) without hysteresis, and (b) with hysteresis included. The M36 is a soft ferromagnetic material with a small hysteresis which can be seen from the figure. While this effect does impact the induced back emf on the bias dc source [6], it does not significantly impact the effective value of the ac inductance over a full fundamental cycle. Therefore, in this study, the hysteresis of this material could be neglected.

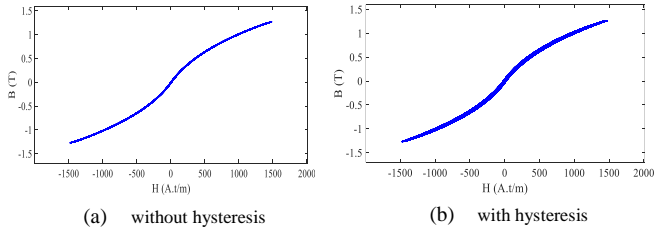

(a) without hysteresis  (b) with hysteresis

Fig.6. B-H characteristics

Figures 7 and 8 illustrate the effects of variation of DBW and dc bias current on the ac inductance of the gapless CVSR, at the rated ac current, in a 3D plot and a 2D projection, respectively. As mentioned previously, theoretically, DBW can assume negative values, meaning that the two ac coils overlap each other partially or completely. Of course, in reality this cannot happen, but it is considered here in order to check the performance of the model. The value of inductance continually decreases as DBW decreases from its highest value of 12″. This happens regardless of the dc current, but the values and the change depend on it. For negative DBWs, when the ac coils overlap, they cancel each other's MMF in the overlapping part due to the opposite current directions. For a complete overlap at -5″, effectively there is no winding and the inductance is zero. This is a result known before hand, and the model is successful in producing the same.

It can be seen from the figures that for small dc bias currents (0-10A), throughout the whole range of DBW, the inductance does not change significantly as the core operates in the linear region of its B-H characteristics. For a dc bias current of 15A, at large DBWs (9.5-12″), the core gets partially saturated within the ac cycle and the inductance decreases compared to its rated (unsaturated) values. At smaller DBWs with smaller rated inductances, the operation remains in the linear region. In these cases, the two ac coils fluxes cancel each other more effectively and the core needs higher dc bias to be forced into saturation. Hence the noticeable drop in the characteristics.

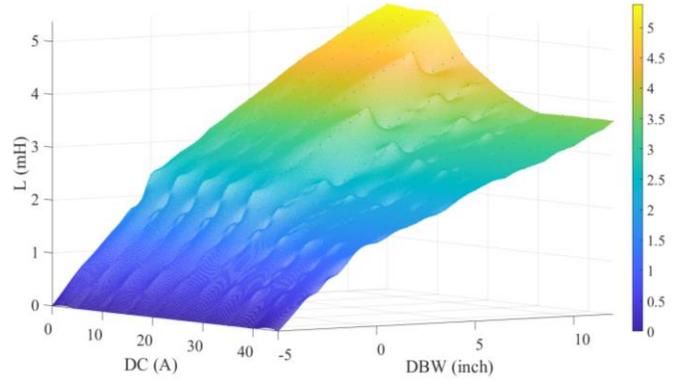

Fig. 7. Inductance vs DBW & dc bias

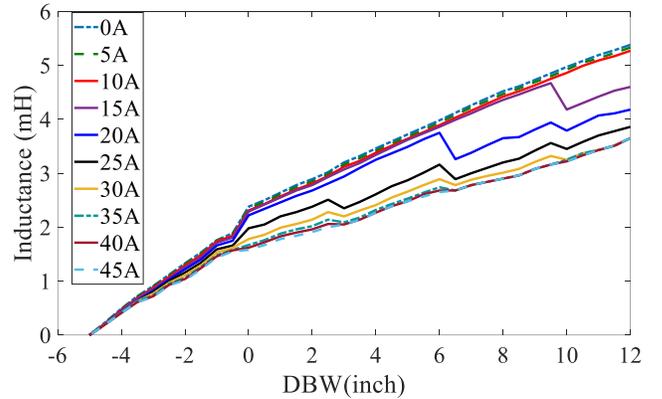

Fig. 8. Inductance vs DBW for different dc bias

For a dc bias current of 20A, at large DBWs (9.5-12″), the core gets further into saturation with further inductance decrease. The partial saturation now starts occurring at smaller DBWs (from 6.5 to 9″) and gets more severe at large DBWs. As the dc bias current increases, the inflection point moves toward smaller DBWs. Eventually, for a dc bias of 45A, the core is fully saturated for all DBWs except the negative ones.

In the negative DBW region, the value of the inductance is low and behaves uniformly for the whole range of dc bias currents. A significantly larger dc current is required to drive the core into the saturation, but this does not have any practical meaning and has not been considered.

Figure 8 shows the effect of partial saturation on the ac inductance of the CVSR at 15 Adc and DBW=10″. At any time, the flux passing through the outer legs is a combination of the ac and dc flux. They add in one part of the outer leg and subtract in the other, as can be seen on Figure 1. If one part of the core operates in partial saturation, the other part is unsaturated and operates in the linear region of the B-H characteristics. In such cases, the inductance is dynamic and changes its value within the ac cycle. The RMS value has been used as a representative single point value.

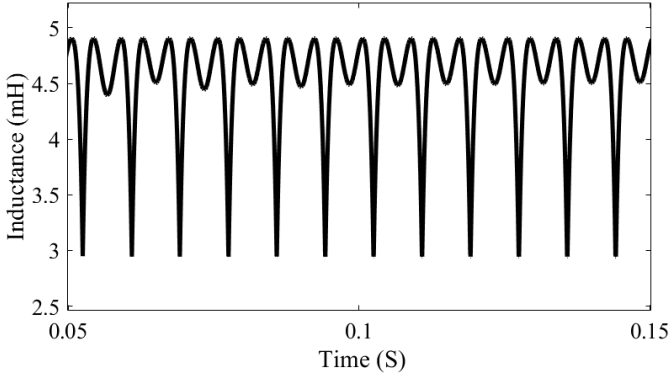

Fig. 8. Dynamic inductance at $I_{dc}$ = 15A and DBW=10″

## APPENDIX

Final expressions for all leakage permeances in the G-C model of the gapless CVSR from Fig. 3 are provided below.

$P_{23-24} = 2(\frac{\mu_0 d_w l_c}{3w_s} + \frac{2\mu_0 d_w}{3\pi} \log(\frac{w_s + \pi r_{mx}}{w_s}))$

$P_{27-28} = \frac{l_c \mu_0 d_s}{w_w^2}(w_w + d_s \log(\frac{d_s}{2w_w}) - \frac{d_s^2}{d_s + 4w_w} + \frac{d_s}{2}) + \frac{4\mu_0 w_w}{3\pi} \log(\frac{d_s + 2\pi d_b}{d_s})$ (inside the winding)

$P_{39-42} = \frac{2 l_c \mu_0 d_s (w_s - w_w)}{(2w_s + d_s)} + \frac{2\mu_0 (w_s - w_w)}{\pi} \log(\frac{d_s + 2\pi d_b}{d_s})$
(outside the winding)

$P_{21-22} = \frac{\mu_0 |DBW| l_c}{w_s} + \frac{\mu_0 |DBW| r_{mx}}{w_s + r_{mx}}$ (virtual air gap permeance)

$r_{mx} = \min(\frac{w_c}{2}, w_o)$

For DBW >0:

$P_{31-32} = 2(\frac{l_e \mu_0}{w_w^2}[\frac{r_1^2}{2} - \frac{d_w r_1}{\pi} + \frac{d_w^2}{\pi^2} \log(\frac{d_w + \pi r_1}{d_w}) + w_w^2 \log(\frac{d_w + \pi r_2}{d_w + \pi r_1}))]$ where: $l_e = l_c + 2w_w$

$r_1 = \min(w_{e1}, w_{e2}, d_w)$
$r_2 = \min(w_{e1}, w_{e2},)$

For DBW <0:

$P_{31-32} = \frac{l_c \mu_0}{256 w_w^2 d_w}[16 k_2^2 + 16\sqrt{2} k_1 k_2^3 + 4 k_1^2 k_2^2 - 2\sqrt{2} k_2 k_1^3 + k_1^4 \ln(1 + 2\sqrt{2} \frac{k_2}{k_1})]$

where: $k_1 = |(w_w - d_w|$
$k_1 = \frac{1}{\sqrt{2}} \min(w_w, 2d_w)$

The exterior adjacent leakage reluctance for dc coils should also include the part of the flux that flows from the side of the yoke as well.

$P_{33-34} = 3(\frac{l_e \mu_0}{w_w'^2}[\frac{r'_1^2}{2} - \frac{d_w' r_1'}{\pi} + \frac{d_w'^2}{\pi^2} \log(\frac{d_w' + \pi r_1'}{d_w'}) + w_w^2 \log(\frac{d_w' + \pi r_2'}{d_w' + \pi r_1'}))]$ where: $l_e = l_c + 2w_w'$

$r_1' = \min(w_{e1}, w_{e2}, d_w')$
$r_2' = \min(w_{e1}, w_{e2},)$

For the other leakages permeances associated with the dc coils, the ac coil parameters should be changed to the dc coil parameters $(d_w', w_w')$.

## VI. CONCLUSION

The paper presents a comprehensive model of a gapless CVSR, by considering all the different leakage fluxes, besides the core characteristics and its nonlinearities. The purpose is to provide a more accurate representation in studying the behavior of the complex hybrid electric-magnetic-electronic device and its interaction with the power system. The G-C approach has been used in modeling the magnetic circuit for correct representation of the interchanged energy between the domains. With this approach, magnetic permeances as energy storage elements are equivalent to capacitors. Simulations with the comprehensive model have been carried out under normal operating conditions, for different values of dc bias current and a whole range of variations of distance between the ac winding coils, a design parameter that defines the nominal, unsaturated inductance. The results obtained reveal some interesting operating characteristics. The ongoing and future work includes studies of this device for various applications in the power systems under different operating conditions.